\documentclass[prd,aps,preprint,nofootinbib,showpacs,superscriptaddress]{revtex4}
\usepackage{graphicx}
\usepackage{hyperref}
\usepackage{amsfonts}
\usepackage{amsmath,amssymb}
\usepackage{bm}
\usepackage{color}
\usepackage{epstopdf}
\usepackage{epsfig}
\usepackage{float}
{\rm }

\def\be{\begin{equation}}
 \def\ee{\end{equation}}
 \def\bea{\begin{eqnarray}}
 \def\eea{\end{eqnarray}}
\def\A{\mathcal{A}}

\def\2{\frac{1}{2}}
\def\4{\frac{1}{4}}

\catcode`\@=11

\def\@normalsize{\@setsize\normalsize{15pt}\xiipt\@xiipt
\abovedisplayskip 14pt plus3pt minus3pt%
\belowdisplayskip \abovedisplayskip
\abovedisplayshortskip  \z@ plus3pt%
\belowdisplayshortskip  7pt plus3.5pt minus0pt}
\def\small{\@setsize\small{13.6pt}\xipt\@xipt
\abovedisplayskip 13pt plus3pt minus3pt%
\belowdisplayskip \abovedisplayskip
\abovedisplayshortskip  \z@ plus3pt%
\belowdisplayshortskip  7pt plus3.5pt minus0pt
\def\@listi{\parsep 4.5pt plus 2pt minus 1pt
            \itemsep \parsep
            \topsep 9pt plus 3pt minus 3pt}}

\def\underline#1{\relax\ifmmode\@@underline#1\else
        $\@@underline{\hbox{#1}}$\relax\fi}
\@twosidetrue \relax

\catcode`@=12

\catcode`\@=11

\def\section{\@startsection{section}{1}{\z@}{3.5ex plus 1ex minus
   .2ex}{2.3ex plus .2ex}{\large\bf}}
\def\ps@headings{\def\@oddfoot{}\def\@evenfoot{}
\def\@oddhead{\hbox{}\hfill
        \makebox[.5\textwidth]{\raggedright\ignorespaces --\thepage{}--
        \hfill }}
\def\@evenhead{\@oddhead}
\def\subsectionmark##1{\markboth{##1}{}}
}

\ps@headings

\catcode`\@=12

\begin{document}

\title{Spontaneously Generated  Inhomogeneous Phases via
Holography
}

\vspace{1.5cm}

\author{James Alsup}
\email{jalsup@umflint.edu}
\affiliation{Computer Science, Engineering and Physics Department,
The University of
Michigan-Flint,
Flint, MI 48502-1907, USA}
\author{Eleftherios Papantonopoulos}
\email{lpapa@central.ntua.gr}
\affiliation{
 Department of Physics, National Technical University of
Athens,
Zografou Campus GR 157 73, Athens, Greece}
\author{George Siopsis}
\email{siopsis@tennessee.edu}
\author{Kubra Yeter}
\email{kyeter@tennessee.edu}
\affiliation{
 Department of Physics and Astronomy, The
University of Tennessee, Knoxville, TN 37996 - 1200, USA
}

\date{\today}
\pacs{11.15.Ex, 11.25.Tq, 74.20.-z}

\vspace{3.5cm}

\begin{abstract}

We discuss a holographic model consisting of a $U(1)$ gauge field
and a scalar field coupled to a charged AdS black hole under a spatially homogeneous chemical potential. By turning on a higher-derivative interaction term between the $U(1)$ gauge field and
the scalar field, a spatially dependent profile of the scalar field is generated spontaneously.
We calculate the critical temperature at which the transition to the inhomogeneous phase occurs for various values of the parameters of the system.
We solve the equations of motion below the critical temperature, and show
that the dual gauge theory on the boundary spontaneously develops a spatially inhomogeneous
charge density.

\end{abstract}

\maketitle

\section{Introduction}

There has been considerable recent activity studying
phenomena at strong coupling using a weakly coupled dual
gravity description. The tool to carry out such a study is the
gauge/gravity duality. This holographic principle
\cite{Maldacena:1997re} has many applications in string theory,
where it is well founded, but it has also been applied to other
physical systems encountered in  condensed matter
physics. One of the most extensively studied condensed matter
systems using the gauge/gravity duality is the holographic
superconductor (for a review see \cite{Horowitz:2010gk}).

The gravity dual of a homogeneous superconductor consists of a
system with a black hole and a charged scalar field. The
black hole admits scalar hair at temperatures lower than a
critical temperature \cite{Gubser:2008px}, while there is no
scalar hair at higher temperatures.  According to the holographic
principle, this breaking of the abelian $U(1)$ symmetry corresponds
in the boundary theory to a scalar  operator which condenses at a
critical temperature dependent on the charge density of the
scalar potential.   The fluctuations of the vector potential give
the frequency dependent conductivity in the boundary theory
\cite{Hartnoll:2008vx}. Backreaction effects on the metric were
 studied in \cite{Hartnoll:2008kx}. In
\cite{Koutsoumbas:2009pa} an exact gravity dual of a gapless
superconductor was discussed in which the charged scalar field
responsible for the condensation was an exact solution of the
equations of motion, and below a critical temperature dressed a vacuum
black hole with scalar hair.

 Apart from holographic applications to conventional homogeneous superconductors,
 extensions  to unconventional superconductors characterized by
 higher critical temperatures, such as cuprates and iron pnictides,
 have also been studied.
 Interesting new features of these  systems
 include competing
orders related to the breaking of lattice symmetries introducing inhomogeneities.  A study of the
effect on the pairing interaction in a weakly
coupled BCS system was performed in \cite{Martin:2005fk}.  Additionally, numerical
studies of Hubbard models \cite{Hellberg:uq,White:kx} suggest that
inhomogeneity might play a role in high-Tc superconductivity.

The recent discovery of transport anomalies in ${\rm La}_{2-x}
{\rm Ba}_x {\rm CuO}_4$ might be explained under the assumption
that the cuprate is a superconductor with a unidirectional charge
density wave, i.e.,  a ``striped" superconductor
\cite{Berg:2009fk}.  Other studies using mean-field theory have
also shown that, unlike the homogeneous superconductor, the striped
superconductor exhibits the existence of a Fermi surface in the
ordered phase \cite{Baruch:2008uq,Radzihovsky:2008kx} and possesses
complex sensitivity to quenched disorder \cite{Berg:2009fk}.
Holographic striped superconductors were discussed in
\cite{Flauger:2010tv} by introducing a modulated chemical potential
producing superconducting stripes below a critical temperature. Properties of the striped
superconductors and backreaction effects were studied in
\cite{Hutasoit:2011rd,Ganguli:2012up}. Striped phases breaking parity and time-reversal
invariance were found in electrically charged AdS-Reissner-Nordstr\"om  black branes with
neutral pseudo-scalars \cite{Donos:2011bh}.  In \cite{Donos:2013bh}, it was shown
that similar phases could be generated in Einstein-Maxwell-dilaton theories
that leave parity and time-reversal invariance intact.

Inhomogeneities also appear in condensed matter systems
other than superconductors. These systems are characterized by
additional ordered states which compete or coexist with
superconductivity \cite{Gabovich,CeCoIn5}. The most important of
these are  charge and spin density waves (CDW and SDW, respectively)
\cite{Gruner}. The development of these states corresponds to
spontaneous modulation of the electronic charge and spin density,
below a critical temperature $T_c$. Density waves are widely
spread among different classes of materials. One may distinguish
between types either orbitally \cite{FISDW},  Zeeman driven
\cite{aperis}, field-induced CDWs, confined \cite{GV}, and even
unconventional density waves \cite{UnconventionalDW}.

The usual approach to study the effect of inhomogeneity at strong
coupling  is to introduce a modulated chemical
potential.   According to the holographic principle this is
translated into a modulated boundary value for the electrostatic
potential in the AdS black hole gravity background. The corresponding
Einstein-Maxwell-scalar systems can be obtained which
below a critical temperature  undergoes a
phase transition to a condensate with a non-vanishing
modulation. Depending on what symmetries are broken, the modulated
condensate gives rise to
ordered states like CDW or SDW
 in the boundary
theory \cite{Aperis:2010cd,Flauger:2010tv}.

To explore  the properties of spatial inhomogeneities in
holographic superfluids, gravitational backgrounds which are not
spatially homogeneous were introduced in
\cite{Maeda:2011pk,Iizuka:2012dk,Liu:2012tr,Donos:2012ra}. In
\cite{Horowitz:2012ky} the breaking of the translational
invariance is sourced by a scalar field with a non-trivial profile
in the $x$-direction. Upon perturbing the one-dimensional
``lattice'', the  Einstein-Maxwell-scalar field equations
were numerically solved at first order and the optical conductivity
was calculated. Further properties of this construction were
studied in \cite{Horowitz:2012gs}.

In this work, we study a holographic superfluid in
which a spatially  inhomogeneous phase is spontaneously generated.
The gravity sector consists of a RN-AdS black hole,
an electromagnetic field, and a scalar field. We
introduce high-derivative interaction terms between the electromagnetic field and
the scalar field.  These higher-order terms are  essential in spontaneously generating
 the inhomogeneous phase in the boundary theory.  Alternative approaches for spontaneously breaking translational symmetries have been found by use of an interaction with the Einstein tensor \cite{Alsup:2012kr,Kuang:2013oqa}, a  Chern-Simons interaction \cite{Donos:2012cs,Nakamura:2009}, and more recently with a dilaton \cite{Donos:2013bh}.

We put the gravitational background on a one-dimensional
``lattice''  generated by an $x$-dependent profile of the
scalar field. At the onset of the condensation of the scalar field,
we calculate the transition temperature. We find that as the
wavenumber of the scalar field increases starting from zero (homogeneous profile), the transition
temperature increases, showing that inhomogeneous configurations dominate at higher temperatures.
We find a
maximum transition temperature corresponding to a certain finite wavenumber.
This is the critical temperature ($T_c$)
of our system.  Below $T_c$ the system undergoes a second order phase
transition to an inhomogeneous phase. This occurs in a range of parameters of the system that we discuss.

We then solve the equations of motion below the critical temperature.
we use perturbation theory to expand the bulk fields right below $T_c$, thus obtaining
an analytic solution to the coupled system of Einstein-Maxwell-scalar field
equations at first order. We find that a spatially inhomogeneous
charge density is spontaneously generated in the boundary theory.

The paper is organized as follows. In section \ref{sec2}, we
present the basic setup of the holographic model, and introduce the higher-derivative couplings. In section
\ref{sec3}, we discuss the instability to a spatially inhomogeneous phase. We calculate numerically the critical temperature of
the system, and analyze its dependence on the various parameters of the system. In section \ref{sec4}, we use perturbation theory to obtain an analytic solution below the
critical temperature, and show that the charge density in the boundary theory is spatially inhomogeneous. Finally, in section
\ref{sec5}, we present our conclusions.

\section{The setup}
\label{sec2}

In this section we introduce a holographic model whose main
feature is the spontaneous generation of spatially inhomogeneous
phases in the boundary theory. This cures the main deficiency of
an earlier proposal \cite{Aperis:2010cd}. This is achieved by
introducing higher-derivative coupling of the electromagnetic field to the scalar
field.

 Consider a system consisting of a $U(1)$ gauge field, $A_\mu$,
 with corresponding field strength $F_{\mu\nu} = \partial_\mu A_\nu -
 \partial_\nu A_\mu$, and a
scalar field $\phi$ with charge $q$ under the $U(1)$ group.
  The fields live in a spacetime of negative cosmological constant $\Lambda = -6/L^2$.

The action is given by
%\begin{equation}
%S = \int d^4 \sqrt{-g} \left[ \frac{R+6/L^2}{16\pi G}
%-\frac{1}{4}F_{\mu\nu}F^{\mu\nu} - \frac{1}{2}\partial_\mu\Psi
%\partial^\mu\Psi - \frac{m^2}{2} \Psi^2 \right] ~.\label{Lag_den}
%\end{equation}
%where $D_\mu\Psi = \partial_\mu \Psi - iqA_\mu\Psi$.
\begin{equation}
S = \int d^4 x \sqrt{-g} \mathcal{L} \ \ , \ \ \ \ \mathcal{L} = \frac{R+6/L^2}{16\pi G}
-\frac{1}{4}F_{\mu\nu}F^{\mu\nu} - (D_\mu\phi)^\ast
D^\mu\phi - m^2 |\phi |^2  ~.\label{Lag_den}
\end{equation}
where $D_\mu\phi = \partial_\mu \phi - iqA_\mu\phi$.
For simplicity, we shall set $16\pi G = L = 1$.

Our main concern is to generate spatially inhomogeneous phases in
the boundary theory. To this end, we may introduce higher-derivative  interaction terms of
the form
\begin{equation}
\mathcal{L}_{\mathrm{int}} = \phi^\ast \left[ \eta \mathcal{G}^{\mu\nu} D_\mu D_\nu + \eta' \mathcal{H}^{\mu\nu\rho\sigma} D_\mu D_\nu D_\rho D_\sigma + \dots \right] \phi + \text{c.c.}  ~,
\end{equation}
which may arise from quantum corrections.
The possible operators in the above expression and their emergence from string theory are worth exploring. Candidates for $\mathcal{G}_{\mu\nu}$ include contributions to the stress-energy tensor form the electromagnetic field and the scalar field, the Einstein tensor  \cite{Alsup:2012kr,Kuang:2013oqa}, etc., and similarly for $\mathcal{H}_{\mu\nu\rho\sigma}$, etc. Here we shall be content with a special choice which leads to inhomogeneities,
\be \mathcal{L}_{\mathrm{int}} = \eta \mathcal{G}^{\mu\nu} (D_\mu \phi)^\ast D_\nu \phi - \eta' | D_\mu \mathcal{G}^{\mu\nu} D_\nu \phi |^2 \ , \label{int_Lag_den}\ee
where
\be \mathcal{G}_{\mu\nu} = T_{\mu\nu}^{\text{(EM)}} +g_{\mu\nu} \mathcal{L}^{\text{(EM)}} = F_{\mu\rho}{F_{\nu}}^{\rho}-\frac{1}{2}g_{\mu\nu}F^{\rho\sigma}F_{\rho\sigma} ~, \label{eq3}\ee
%\begin{equation}
%S_{\mathrm{int}} =\frac{\eta}{2} \int d^4 x \sqrt{-g} \left[ \left(
%G^{\mu\nu}-3g^{\mu\nu}\right) (D_\mu \phi)^\ast D_\nu
%\left\{ \mathcal{J} (- \alpha
%\mathcal{D}_\sigma\mathcal{D}^\sigma)\phi\right\} +c.c. \right] ~,
%\label{int_Lag_den}
%\end{equation}
coupling the scalar field $\phi$ to the
% Einstein tensor
%\begin{equation}
%G_{\mu\nu} = R_{\mu\nu} - \frac{1}{2} R g_{\mu\nu}~,
%\end{equation}
%as well as the
gauge field.
This coupling is the essential tool for the generation of spatial inhomogeneities.
From the action \eqref{Lag_den}, together with the interaction term
\eqref{int_Lag_den}, we obtain the Einstein equations
\begin{equation}
R_{\mu\nu} - \frac{1}{2} R g_{\mu\nu} - 3 g_{\mu\nu} = \frac{1}{2} T_{\mu\nu}
~,\label{Eineq}
\end{equation}
where $T_{\mu\nu}$ is the stress-energy tensor,
%\be \nabla^\mu T_{\mu\nu} = 0 \ee
\begin{equation}
T_{\mu\nu} = T_{\mu\nu}^{(EM)} + T_{\mu\nu}^{(\phi)} + \eta\Theta_{\mu\nu} + \eta' \Theta_{\mu\nu}'~,
\end{equation}
containing a gauge, scalar, and interaction term contributions, respectively,
%. Assuming that the effect of the second interaction term (with coupling constant $\eta'$) is negligible beyond determining the size of the inhomogeneity (lattice spacing), to be confirmed by the results, we obtain the explicit expressions
%\bea T_{\mu\nu}^{(EM)} &=&F_{\mu\rho}{F_{\nu}}^{\rho}-\frac{1}{4}g_{\mu\nu}F^{\rho\sigma}F_{\rho\sigma}~,  \nonumber\\
%T_{\mu \nu}^{(\Psi)} &=& \nabla_\mu\Psi \nabla_\nu \Psi - \frac{1}{2}
%g_{\mu\nu} \nabla_\alpha \Psi \nabla^\alpha \Psi - m^2 g_{\mu\nu}
%\Psi^2~, \nonumber\\
% \Theta_{\mu \nu}&=& - \frac{1}{2} (R +12) \nabla_\mu\Psi
%\nabla_\nu\Psi
%+ R^\alpha_\nu \nabla_\alpha\Psi \nabla_\mu \Psi + %R^\alpha_\mu \nabla_\alpha\Psi \nabla_\nu
%\Psi - \frac{1}{2} G_{\mu\nu} \nabla_\alpha\Psi \nabla^\alpha\Psi
%\nonumber
%\\&+& R_{\mu\alpha\nu\beta} \nabla^\alpha\Psi \nabla^\beta \Psi + \nabla_\mu \nabla^\alpha \Psi
% \nabla_\nu \nabla_\alpha \Psi - \nabla_\mu\nabla_\nu \Psi \Box
% \Psi\no-number
%\\&+& g_{\mu\nu} \left[ -\frac{1}{2} \nabla^\alpha\nabla^\beta
%\Psi \nabla_\alpha\nabla_\beta\Psi + \frac{1}{2} (\Box\Psi)^2 -
%\left( R^{\alpha\beta} -3 g^{\alpha\beta} \right)
%\nabla_\alpha\Psi \nabla_\beta \Psi \right]~. \eea
\bea T_{\mu\nu}^{(EM)} &=&F_{\mu\rho}{F_{\nu}}^{\rho}-\frac{1}{4}g_{\mu\nu}F^{\rho\sigma}F_{\rho\sigma}~,  \nonumber\\
T_{\mu \nu}^{(\phi)} &=& (D_\mu\phi)^\ast D_\nu \phi +  D_\mu \phi (D_\nu\phi)^\ast  -
g_{\mu\nu} (D_\alpha \phi)^\ast D^\alpha \phi - m^2 g_{\mu\nu}
|\phi|^2~, \nonumber\\
 \Theta_{\mu \nu}&=& \frac{2}{\sqrt{-g}}\frac{\delta}{\delta g^{\mu\nu}}\int d^4 x\sqrt{-g} \mathcal{G}^{\mu\nu} (D_\mu\phi)^\ast D_\nu \phi~,
\nonumber\\
 \Theta_{\mu \nu}'&=&-\frac{2}{\sqrt{-g}}\frac{\delta}{\delta g^{\mu\nu}}\int d^4 x\sqrt{-g}| D_\mu \mathcal{G}^{\mu\nu} D_\nu \phi |^2 ~. \eea
 Varying the
Lagrangian with respect to $A_\mu$ we find the Maxwell equations
% (with $\eta'=0$),
\be\label{maxeq}
\nabla_{\mu}F^{\mu \nu}= J^\nu \ , \ee
where $J_\mu$ is the current,
\be J_\mu = q J_\mu^{(\phi)} + \eta \mathcal{J}_\mu +  \eta' \mathcal{J}_\mu' ~, \ee
containing scalar and interaction term contributions, respectively,
\bea J_\mu &=& i  \left[ \phi^\ast D_\mu
\phi -(D_\mu \phi)^\ast \phi\right] ~, \nonumber\\
\mathcal{J}_\mu &=& \frac{1}{\sqrt{-g}}\frac{\delta}{\delta A^\mu}  \int d^4 x\sqrt{-g} \mathcal{G}^{\mu\nu} (D_\mu\phi)^\ast D_\nu \phi ~,
\nonumber\\
\mathcal{J}_\mu' &=& - \frac{1}{\sqrt{-g}}\frac{\delta}{\delta A^\mu} \int d^4 x\sqrt{-g}| D_\mu \mathcal{G}^{\mu\nu} D_\nu \phi |^2 ~. \eea
%We can use gauge invariance to set the St\"uckelberg field
%\[ \omega = 0~. \]
%
Finally, the equation of motion for the scalar field is
\bea
D_\mu D^\mu \phi-m^2 \phi=\eta  D_\mu \left(\mathcal{G}^{\mu\nu}D_\nu \phi\right)+\eta' D_\rho (\mathcal{G}^{\mu\rho}D_\mu (D_\nu (\mathcal{G}^{\nu\sigma} D_\sigma \phi)))
%\nonumber\\ &+& {\bf (q-term)}
. \label{waveeq}
\eea
%where we used the fact that the Einstein tensor is conserved ($\nabla_\mu G^{\mu\nu} = 0$) and that our interest is in solutions with $\mathcal J \phi = J \phi$ for scalar function $J$.
%It is important to include the term proportional to $\eta'$ in the scalar equation because its solution will determine the lattice spacing.
Our aim is to study the Einstein-Maxwell-scalar system of
equations first at the critical temperature, and then below the
critical temperature using perturbation theory.

\section{The critical temperature}
\label{sec3}

At the critical temperature, we have $\phi=0$. The
Einstein-Maxwell system has a static solution with metric of the
form
\be ds^2 = \frac{1}{z^2} \left[ - h(z) dt^2 + \frac{dz^2}{h(z)} + dx^2 + dy^2\right]~. \ee
The system possesses a scaling symmetry. The arbitrary scale is often taken to be the radius of the horizon. It is convenient to fix the scale by using a radial coordinate $z$ so
the horizon is at $z=1$. Since the scale has been fixed, we should only be reporting on scale-invariant quantities.

The Maxwell equations admit the solution
\be\label{eq2_11a} A_t = \mu (1-z) ~, \ee
so that the $U(1)$ gauge field has an electric field in the
$z$-direction equal to the chemical potential, $E_z = \mu$.

The Einstein equations are then solved by
\be\label{eq2_11b} h(z) = 1 - \left( 1 + \frac{\mu^2}{4} \right) z^3 + \frac{\mu^2}{4} z^4~. \ee
The temperature is given as
\bea
\frac{T_c}{\mu} = - \frac{h'(1)}{4\pi\mu}  = \frac{3}{4\pi\mu} \left( 1 - \frac{\mu^2}{12} \right)~,
\label{temper}
\eea
where we divided by $\mu$ to create a scale-invariant quantity.

Additionally, at the critical temperature the scalar field satisfies the wave equation,
\begin{equation}
\partial_z^2 \phi+\left[\frac{h'}{h}-\frac{2}{z}\right]\partial_z \phi+\frac{1}{h}\left(1-\eta \mu^2 z^4 -\eta' \mu^4z^{10} \nabla_2^2\right)\nabla_2^2 \phi-\frac{1}{h}\left[\frac{m^2}{z^2}-q^2 \frac{A_t^2}{h}\right]\phi=0~, \label{weqPsi}
\end{equation}
where $\nabla_2^2 = \partial_x^2 + \partial_y^2$, and we fixed the gauge so that $\phi$ is real.
%F(J) &=& \frac{\eta}{ 2z^4} \left(G^{\mu\nu}-3g^{\mu\nu}\right) \nabla_\mu\nabla_\nu  J~.

The wave equation \eqref{weqPsi} can be solved by separating
variables, \bea \phi(z,x,y)=\Phi(z) Y(x,y)~, \label{ansatzphi} \eea
where $Y$ is an eigenfunction of the two-dimensional Laplacian,
\be \nabla_2^2 Y = - \tau Y~. \ee We will keep translation
invariance in the $y$-direction, and concentrate on the one-dimensional
``lattice" defined by
 \be\label{eqlat} Y =\cos (k x)~, \ee with $\tau = k^2$, and leave the  two-dimensional lattices
 for future study.

\begin{figure}
\begin{center}
\includegraphics[width=0.8\textwidth]{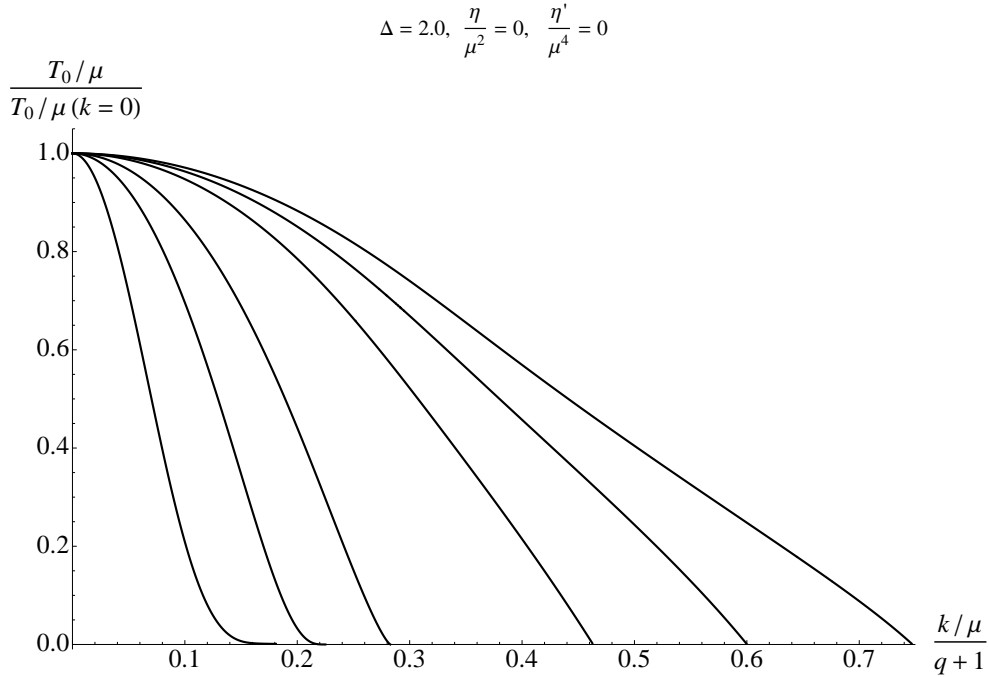}
\end{center}
\caption { \label{temperature_q1} Dependence of the transition temperature $T_0/\mu$
  on the wavenumber $k/\mu$ for $\Delta=2$ in the absence of higher-derivative interactions ($\frac{\eta}{\mu^2}=\frac{\eta'}{\mu^4} = 0$).  From left-to-right the lines correspond to $q=0, 1, 1.5, 3, 5, 10$.  The maximum transition temperature (i.e., the critical temperature) occurs at $k=0$ and the homogeneous configuration is dominant.
}
\end{figure}

The radial function $\Phi(z)$ satisfies the wave equation
\bea
%\Phi''+\left[\frac{h'}{h}+\frac{f_+'}{f_+}-\frac{2}{z}\right] \Phi'-\frac{\tau}{h} \frac{f_-}{f_+} \Phi-\frac{m^2}{z^2h f_+} \Phi + %\frac{q^2 A_t^2}{h^2} \Phi \nonumber\\
%-3 \alpha  \mu ^2 \tau z^5\left[\left(\frac{h'}{h}+\frac{3}{z}\right)\left(1-\frac{1}{f_+}\right)+\frac{f_+'}{f_+}\right] \Phi=0
\Phi''+\left[\frac{h'}{h}-\frac{2}{z}\right]\Phi'+\frac{\tau}{h}\left[1-\eta \mu^2 z^4-\eta' \tau \mu^4  z^{10} \right] \Phi-\frac{1}{h}\left[\frac{m^2}{z^2}-q^2 \frac{A_t^2}{h}\right]\Phi=0~.
\eea
%and we have used the notation $\partial_z \equiv \, '$.
The asymptotic behavior (as $z\to 0$) is $\Phi \sim z^\Delta$,
where $\Delta (\Delta -3) = m^2$. It is convenient to write
\be \Phi (z) = \frac{\langle\mathcal{O}_\Delta \rangle}{\sqrt{2}} z^\Delta F(z) \ \ , \ \ \ \ F(0) =1~. \label{Fscalar}\ee
For general $\Delta$, we obtain
\begin{equation}
F''+\left[\frac{2(\Delta-1)}{z}+ \frac{h'}{h}\right]F'+
\Big[ -\frac{\tau \left( 1 - \eta\mu^2 z^4 -\eta' \tau \mu^4 z^{10} \right)}{h} + \frac{q^2 A_t^2}{ h^2}+\frac{m^2}{z^2}\left(1-
\frac{1}{ h} \right)
+\frac{\Delta h'}{zh} \Big ] F=0~.
\label{critical}
\end{equation}
%where we have used the notation $\partial_z \equiv \, '$.
The maximum transition temperature of the system can be calculated
by solving \eqref{critical} numerically, and using the expression
\eqref{temper} for the temperature. The maximum transition temperature is
the critical temperature $T_c$ of the system.  Figure \ref{temperature_q1} shows the numerically calculated transition temperature without the higher-derivative
%Einstein
couplings ($\frac{\eta}{\mu^2}=\frac{\eta'}{\mu^4} =0$) dependent on the wavenumber $k$.  Both quantities are divided by the chemical potential $\mu$ to render them dimensionless.    The maximum value is found  at $k=0$, which shows that the homogeneous solution is dominant.
When the higher-derivative  interaction terms are turned on, the critical temperature
$T_c$ of the system also depends on the coupling constants $\eta$, $\eta'$, and the homogeneous solution no longer dominates.

In the limit of vanishing second higher-derivative coupling, $\eta' \to 0$, we may analytically calculate the asymptotic critical temperature.  The latter is found in the limit in which the wavenumber diverges ($\tau\to \infty$). In this limit, the wave equation \eqref{critical} is  dominated by the term proportional to $\tau$ near the horizon ($z \to 1$), %realizing the constraint $f_{-}(1) = 0$.
thus giving the critical value for the chemical potential as $\mu_c^4 = \frac{\mu^2}{\eta}$, and corresponding temperature \eqref{temper}
\be\label{asyT}
\lim_{\tau/\mu^2 \to \infty} \frac{T_0}{\mu} = \frac{3}{4\pi} \left( \frac{\eta}{\mu^2} \right)^{1/4} \left(1 - \frac{1}{12 \sqrt{\frac{\eta}{\mu^2}} } \right)~.
\ee
The critical temperature for a standard Einstein-Maxwell-scalar system, i.e., $\frac{\eta}{\mu^2} = \frac{\eta'}{\mu^4}=0$, with a neutral scalar was calculated in \cite{Alsup:2011lt}.   For $\Delta=2$, it was found that $\frac{T_c}{\mu} \approx .00009$.  For $\eta$ large enough, the asymptotic ($\tau/\mu^2 \to \infty$) transition temperature will be higher than that of the homogeneous solution. In this case, the transition temperature monotonically increases as we increase the wavenumber $k$ and asymptotes to \eqref{asyT}.
Hence the
%Einstein
higher-derivative coupling's encoding of the electric field's back reaction near the horizon is the cause of spontaneous generation of spatial modulation.

As we switch on $\eta' >0$, the transition temperature is bounded from above by \eqref{asyT}. It attains a maximum value at a finite $k$. Thus the second higher-derivative coupling acts as a UV cutoff on the wavenumber, which in turn determines the size of the ``lattice" through $k = \frac{2\pi}{a}$, where $a$ is the lattice spacing. As $\eta' \to 0$, the lattice spacing also vanishes ($a\to 0$) and the wavenumber diverges ($k\to\infty$). We will focus on small-to-zero charge, realizing a transition temperature at zero wavenumber \emph{below} \eqref{asyT}, which guaranties that $k\ne 0$ at the maximum transition temperature, hence the dominance of inhomogeneous modes.  For our purposes, we do not expect any quantitative differences between small and zero charge, even with a neutral scalar leaving the boundary $U(1)$ symmetry intact.

\begin{figure}
\begin{center}
\includegraphics[width=0.8\textwidth]{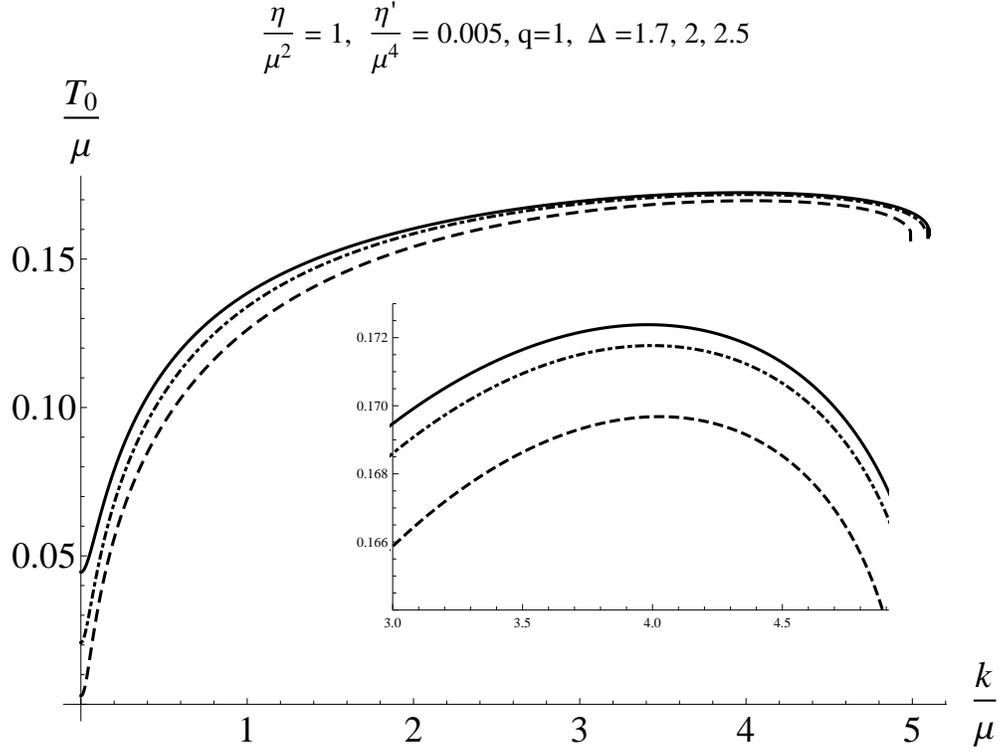}
\end{center}
\caption {Dependence of the transition temperature $\frac{T_0}{\mu}$
  on the wavenumber $\frac{k}{\mu}$ for $\frac{\eta}{\mu^2}=1$, $\frac{\eta'}{\mu^4}=0.005$, $q=1$, and $\Delta=1.7$ (solid line), $2$ (dash-dotted line), and $2.5$ (dotted line).
} \label{temperature_first}
\end{figure}

\begin{figure}
\begin{center}
{\includegraphics[width=.45\textwidth]{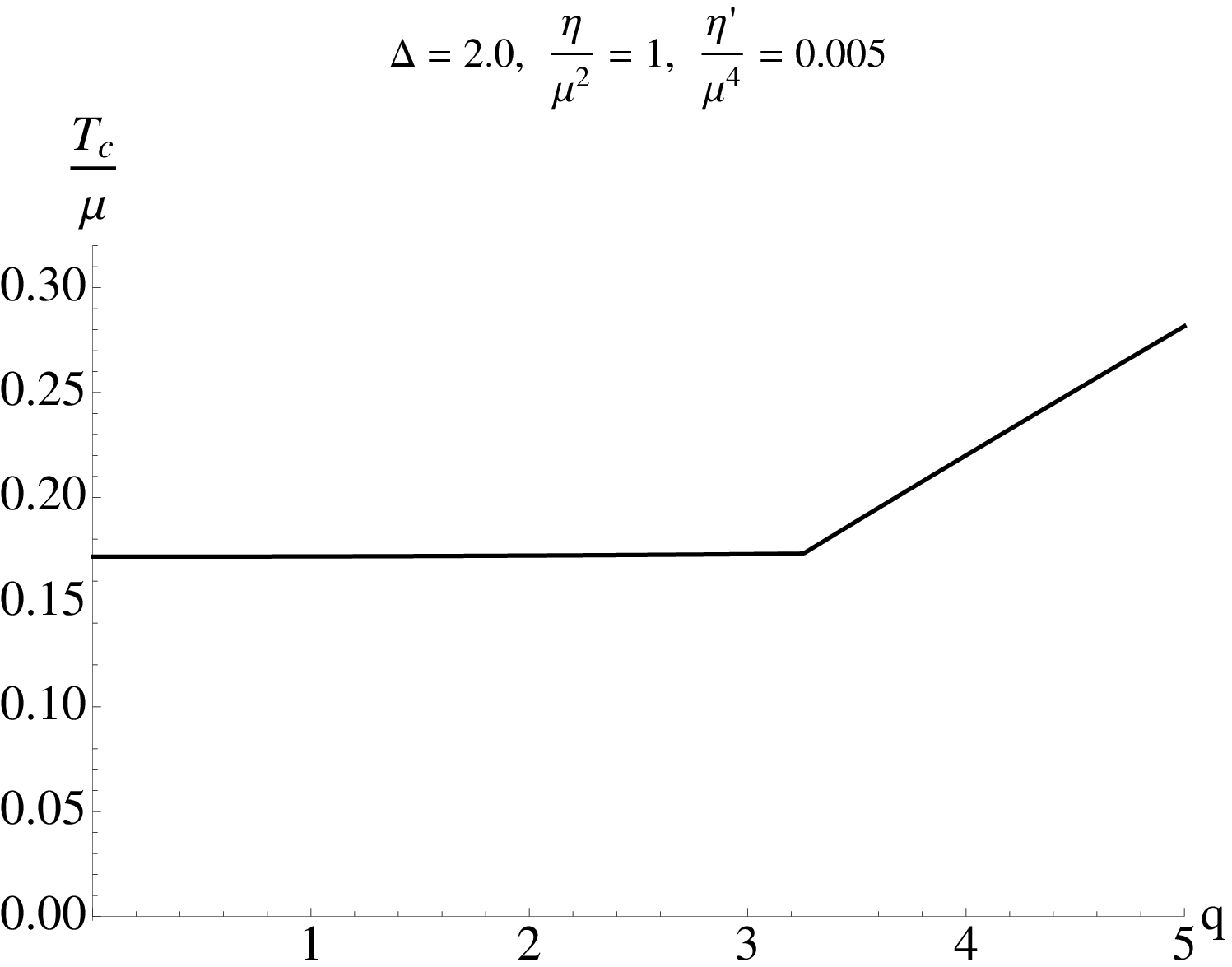}}
%\scalebox{0.7}{\includegraphics{eta1alpha0p0007delta1delta1p7muvsalpha.eps}}
$\quad$ {\includegraphics[width=.45\textwidth]{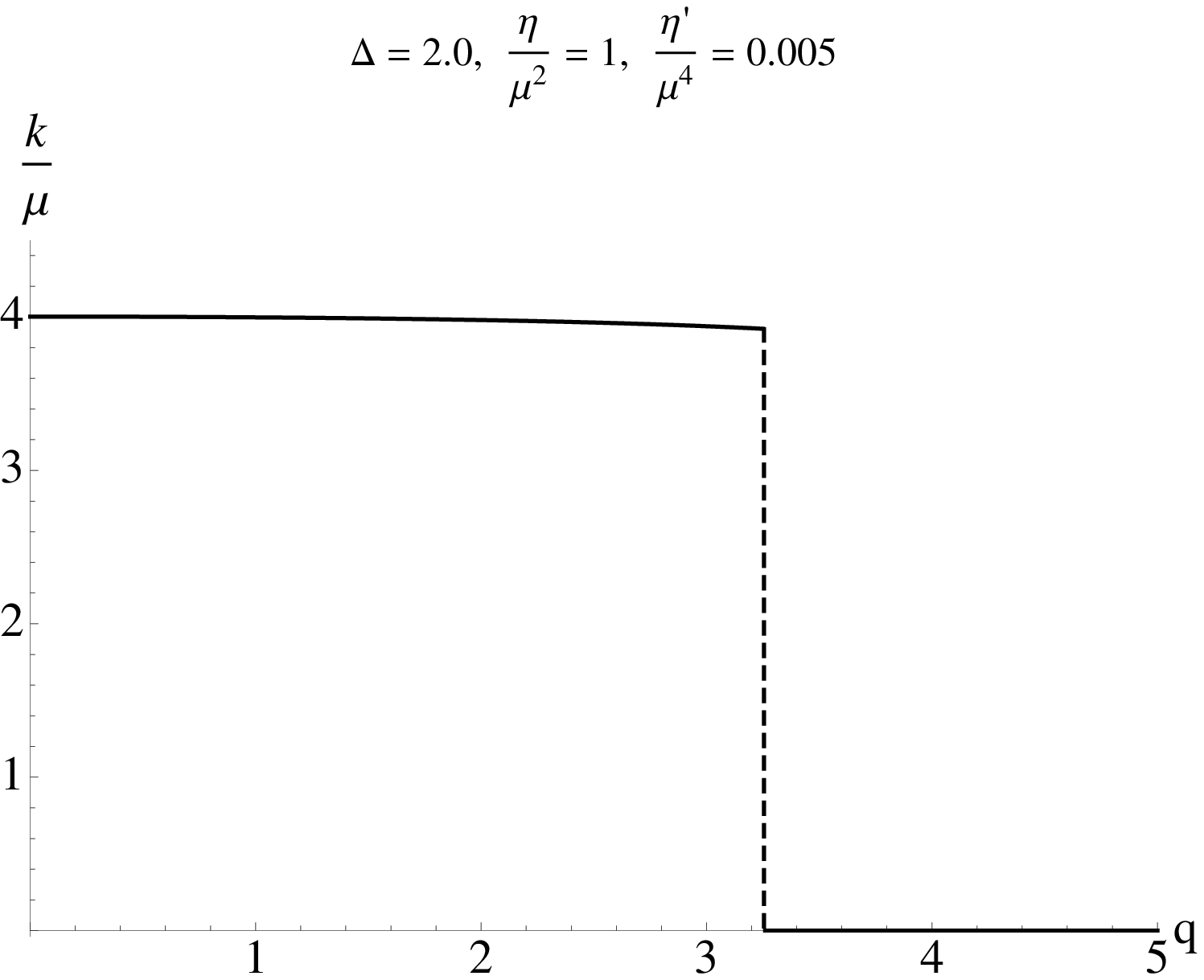}}
\end{center}
\caption {The critical temperature $\frac{T_c}{\mu}$ (left panel) and corresponding wavenumber $\frac{k}{\mu}$ (right panel) as functions of $q$ with
  $\frac{\eta}{\mu^2}=1$, $\frac{\eta'}{\mu^4}=0.005$, and $\Delta=2$.  When $q\lesssim 3.2$,  the charge is small enough that the higher-derivative couplings spontaneously generate inhomogeneity.  Above that range, the homogeneous scalar is dominant.
} \label{temperature_q2}
\end{figure}

In Figure \ref{temperature_first}, we show the transition
temperature $T_0$ as a function of the wavenumber $k$ with $q=1$, $\frac{\eta}{\mu^2} = 1$, $\frac{\eta'}{\mu^4} = 0.005$, for various values of the conformal dimension of the scalar field $\Delta$.
The critical temperature of the system is the maximum transition temperature which occurs at finite $k/\mu\approx 3.98$.
The effect of the charge $q$ is shown in Fig. \ref{temperature_q2}.  It can be seen that for finite
%Einstein
coupling constants $\eta$ and $\eta'$, and small enough charge $q$, the system produces a {\it critical} temperature at non-vanishing finite values of $k/\mu$.  For large enough $q$, the homogeneous solution ($k=0$) remains the dominant solution.

\begin{figure}
\begin{center}
{\includegraphics[width=.45\textwidth]{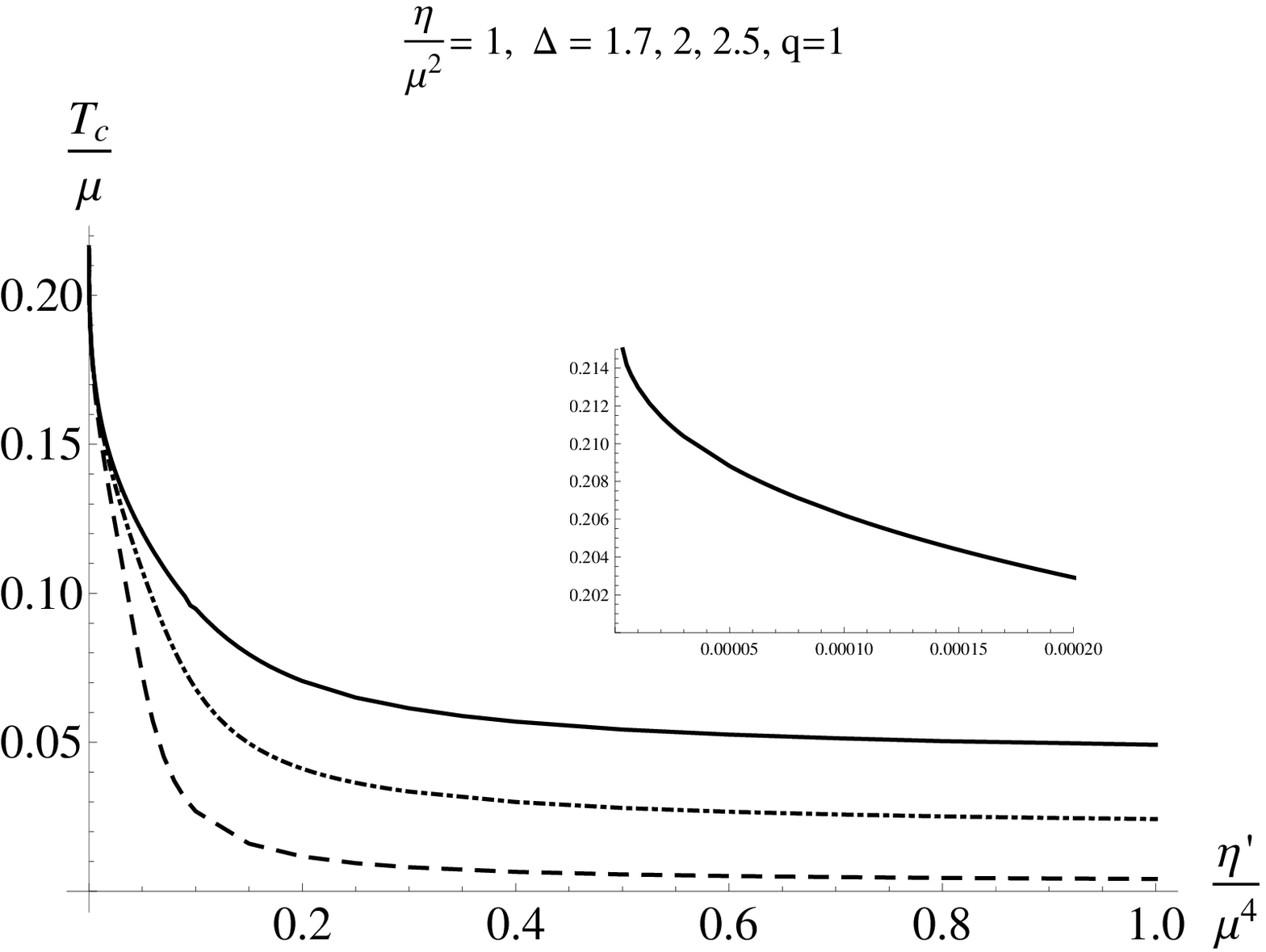}}
%\scalebox{0.7}{\includegraphics{eta1alpha0p0007delta1delta1p7muvsalpha.eps}}
{\includegraphics[width=.45\textwidth]{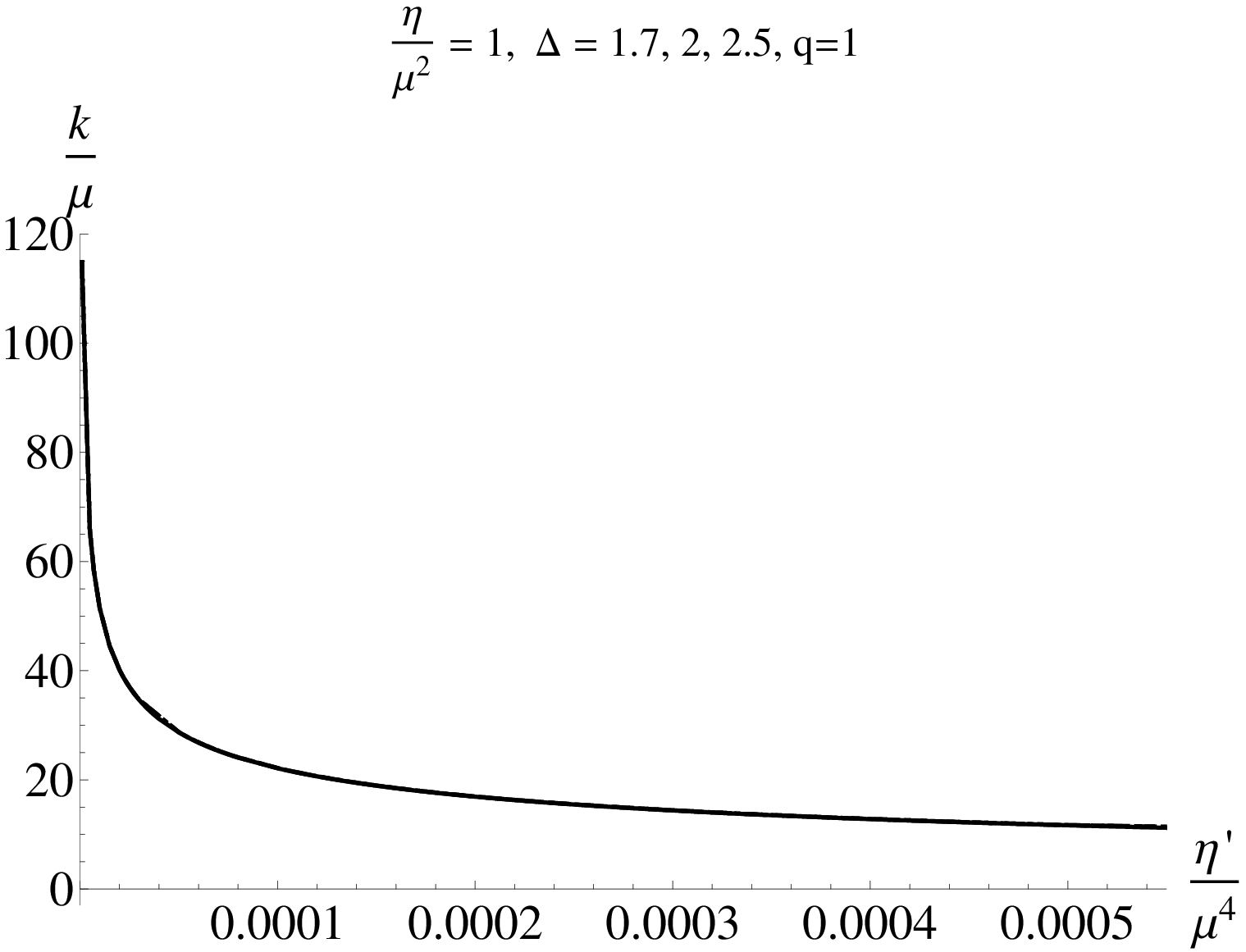}}
\end{center}
\caption {Dependence of the critical temperature $\frac{T_c}{\mu}$ (left panel) and corresponding wavenumber
$\frac{k}{\mu}$ (right panel) on the second higher-derivative coupling constant $\frac{\eta'}{\mu^4}$, for $\frac{\eta}{\mu^2}=1$, $q=1$, $\Delta=1.7$ (solid), $2$ (dash-dotted), and $2.5$ (dashed).
The inset is shown only for $\Delta=2$ because, at the scale shown, no difference can be seen between the three different conformal dimensions of the full graph.
} \label{temperature_second}
\end{figure}

 Figure
\ref{temperature_second} shows the dependence of the
critical temperature $T_c$ and wavenumber $k$ on the second higher-derivative coupling constant $\frac{\eta'}{\mu^4}$.  There is little to no dependence of $T_c$ on the conformal dimension $\Delta$.
We see numerical confirmation that in the absence
of the second coupling ($\eta' =0$), the maximum transition
temperature corresponds to $k \to \infty$. Thus $\eta'$ acts as an effective UV cutoff, determining the wavenumber $k$ at the critical temperature, and therefore the size of the ``lattice" of the system (if $k = \frac{2\pi}{a}$, where $a$ is the lattice spacing). The value of the wavenumber decreases with increasing coupling $\eta'$, as shown on the right panel of Fig.\ \ref{temperature_second}.  The UV cutoff $\eta'$, or effective lattice spacing, can be understood as stabilizing the inhomogeneous modes introduced by the
%Einstein
first higher-derivative coupling $\eta$.  Looking forward, we trust our linearization below the critical temperature because it will not rely on a gradient expansion but on an order parameter proportional to $\left(T-T_c\right)^{1/2}$.

%\begin{figure}
%\begin{center}
%\includegraphics[scale=0.8]{TcvsBxi10delta1delta2.eps}
%\includegraphics[scale=0.8]{kvsBxi10delta1delta2.eps}
%\end{center}
%\caption {Dependence of critical temperature $T_c$ (left panel)
%and dependence of wave vector $k$ (right panel) on the change of
%$\mathcal{B}$ for $\Delta=1$ and $\Delta=2$. } \label{delta1}
%\end{figure}

%\begin{figure}
%\begin{center}
%\includegraphics[scale=0.8]{xi10delta2p5delta2delta1.eps}
%\includegraphics[scale=0.8]{xi10delta2p5delta2delta1kvsB.eps}
%\end{center}
%\caption {Dependence of critical temperature $T_c$ (left panel)
%and dependence of wave vector $k$ (right panel) on the change of
%$\mathcal{B}$ for $\Delta=1$, $\Delta=2$, $\Delta=2,5$ and
%$\xi=10$. } \label{zerotem}
%\end{figure}

Summarizing, at the critical temperature the electric field
backreacts on the system, the Einstein-Maxwell field equations
admit solutions with a  spatially dependent scalar field while the
electric field attains a constant value equal to the
chemical potential. In the next section we will
perturb around the critical temperature and show the system
develops a spatially inhomogeneous phase in the boundary theory.

%Due to the fact that the UV cutoff parameter $\alpha$ contributes
%in higher order terms we will continue our calculations with the
%original Lagrangian density terms (Eqs.\
%(\ref{Lag_den})-(\ref{int_Lag_den})) below the critical
%temperature.\\
%
%{\bf What do you mean by that? You consider an action without the cutoff $\alpha$?}\\

\section{Below the critical temperature}
\label{sec4}

In this section we study the system below the critical
temperature.
%For simplicity, we will focus on a neutral scalar ($q=0$).  The effects of a finite $q$ will not qualitatively alter the results.
 The equations of motion resulting from the
considered action \eqref{Lag_den} together with \eqref{int_Lag_den} may be perturbed near the
critical temperature with spatially dependent solutions. We
will study the behavior of the system analytically, leaving a full numerical study for the future.
To simplify the discussion somewhat, we shall assume that the effects of the cutoff are negligible
%, i.e., $\mathcal{J} \approx 1$
near the critical temperature ($T\approx T_c$), and set $\eta'=0$.  We will build a perturbative expansion on the departure below $T_c$ and not in terms of gradients or momentum of the scalar mode.
It is straightforward, albeit tedious, to include the effects of the cutoff below $T_c$.

Below the
critical temperature the scalar field backreacts on the metric. Consider the following \emph{ansatz}
\be ds^2 = \frac{1}{z^2} \left[ - h(z,x)e^{-\alpha(z,x)} dt^2 + \frac{dz^2}{h(z,x)} +e^{\beta(z,x)} dx^2 +e^{-\beta(z,x)} dy^2\right]~.
\label{ansatz}
\ee
%\subsection{Perturbations around the critical temperature \label{subsec2}}
To solve the equations of motion \eqref{Eineq}, \eqref{maxeq}, and
\eqref{waveeq} below the critical temperature $T_c$, we
expand in the order parameter \be \xi = \frac{\langle
\mathcal{O}_\Delta \rangle}{\sqrt{2}}~, \label{condensate1} \ee
and write
\bea
%\begin{split}
h(z,x)&=& h_0(z)+\xi^2 h_1(z,x) + \mathcal{O} (\xi^4)~,\nonumber \\
\alpha(z,x)&=&\xi^2 \alpha_1(z,x)  + \mathcal{O} (\xi^4)~,\nonumber \\
\beta(z,x)&=&\xi^2 \beta_1(z,x) + \mathcal{O} (\xi^4)~, \nonumber \\
\phi(z,x)&=&\xi \phi_0(z,x)+\xi^3 \phi_1(z,x) + \mathcal{O} (\xi^5)~, \nonumber
\\A_t(z,x)&=&A_{t0}(z)+\xi^2 A_{t1}(z,x) +\mathcal{O}(\xi^4)~, \label{pertscalar}
%\end{split}
\eea
where $A_{t0}$,
 $h_{0}$, and $\xi\phi_{0}$  are  defined
at the critical temperature $T_c$ by eqs.\ \eqref{eq2_11a}, \eqref{eq2_11b}, and \eqref{weqPsi}, respectively. The chemical potential is given as
\be\label{eq2_23}
\mu \equiv A_t (0,x) = \mu_0 + \xi^2 \mu_1 + \mathcal{O} (\xi^2) \
\ , \ \ \ \ \mu_0 = A_{t0}(0) \ \ ,  \ \ \ \ \mu_1 = A_{t1}(0,x)~. \ee
It should be noted that we are working with an ensemble of fixed chemical potential, which seems to contradict eq.\ \eqref{eq2_23} in which the chemical potential appears to receive corrections below the critical temperature. However, the reported chemical potential is measured in units in which the radius of the horizon is $1$ and a change in $\mu$, in these units, is due to a change in our scale as we lower the temperature.

At each given order of the parameter $\xi$, only a finite
number of modes of the various fields are generated. At
$\mathcal{O} (\xi^2)$, we have only $0$ and $2k$ Fourier modes,
%After substituting perturbations
%(\ref{pertscalar})  in the Maxwell's (\ref{maxeq}) and Einstein's (\ref{Eineq}) equations we also make a Fourier series expansion.
 \bea
%\begin{split}
h_1(z,x)&=&z^3 \left( h_{10}(z)+ h_0^2(z) h_{11}(z) \cos 2kx\right) ~,\nonumber \\
\alpha_1(z,x)&=& \alpha_{10}(z)+ z^3h_0(z) \alpha_{11}(z) \cos 2 kx ~,\nonumber \\
\beta_1(z,x)&=&\beta_{10}(z)+z^3 \beta_{11}(z) \cos 2 kx ~, \nonumber
\\A_{t1}(z,x)&=& A_{t10}(z)+ zh_0(z)A_{t11}(z) \cos 2kx~, \label{fourier}
%\end{split}
\eea where we included explicit factors of $z$ and $h_0(z)$ for
convenience. From \eqref{eq2_23}, we obtain the boundary condition
\be\label{eq2_25} A_{t10} (0) =\mu_1~. \ee
Then from the Maxwell equation \eqref{maxeq}, and the boundary
condition \eqref{eq2_25}, we find
\be\label{eqV10}
A_{t10}(z) = C(1-z)+\frac{\mu_0}{4} \int_1^z dw\, \int_1^w dw' {w'}^{2\Delta-2} h_0(w') \mathcal{A}(w')
  ~,\ee
where
\bea\label{eqC} C &=& \mu_1 + \frac{\mu_0}{4} \int_0^1 dz\,\int_1^z dw {w}^{2\Delta-2} h_0(w) \mathcal{A}(w)~, \nonumber\\
\mathcal{A} (z) &= & \left[ q^2 \frac{\mu_0^2(1-z)^2 z^3 +4 q^2 (1-z) h_0(z)}{h_0^2(z)}+z\left(\Delta^2+8 k^2 \eta(1+\Delta) z^2\right)  \right] F^2(z)\nonumber\\
&& +2 z^2 \left[\Delta+4 k^2\eta  z^2 \right]  F(z) F'(z)+z^3  [F'(z)]^2~.
\eea
%and $\Phi_0$ comes from the zeroth order of (\ref{ansatz}).
Thus the integration constant $\mathcal{C}$ is expressed in terms of the chemical potential parameters $\mu_0$ and $\mu_1$. While $\mu_0$ is determined at the critical temperature, $\mu_1$ still needs to be determined. Subsequently, we will determine $\mathcal{C}$ using the scalar equation and use that value in eq.\ \eqref{eqV10} to find $\mu_1$.

After some algebra, from the
Einstein equations we deduce that
the mode function $\alpha_{10}(z)$ is given by
\be \alpha_{10} (z) = \frac{1}{2} \int_0^z dw\, w^{2\Delta -1} \left[ \left( q^2 \mu_0^2 \frac{(1-w)^2 w^2}{h_0^2(w)} +\Delta ^2 \right) F^2(w)+2 \Delta  wF(w) F'(w)+w^2 [F'(w)]^2\right]\ .\label{eq4} \ee
%where
%\bea
%\hat\alpha_{10}(z)&=& [\Phi_0'(z)]^2 + \eta \Big[ -k^2 \Phi_0^2(z) -2\left( 1+ \frac{k^2}{2} z^2 -5(1+\mu^2/4) z^3 + %\frac{5\mu^2}{4} z^4 \right) [\Phi_0'(z)]^2 \nonumber\\
%&&  -2z h_0(z) \Phi_0'(z)\Phi_0''(z) -4k^2z \Phi_0(z)\Phi_0'(z) - k^2 z^2 \Phi_0(z) \Phi_0''(z) \Big] \, . \label{eq4h}
%\eea
Notice that the mode function  $\alpha_{10}$ contributes to $\alpha_1$ at an order higher than
$\mathcal{O} (z^3)$ near the boundary for $\Delta > \frac{3}{2}$.

%and $\Phi_0$ comes from the zeroth order of (\ref{ansatz}).

The mode function $\beta_{10}(z)$ is given by
\begin{equation}
\beta_{10}(z)= \frac{k^2}{2} \int_0^z \frac{ dw\, {w}^2
}{h_0(w)}\int_{w}^1 dw'\, {w'}^{2\Delta-2} (1-\eta \, \mu_0^2 \, {w'}^4) F^2(w')
\label{eq5}~,\ee
%with
%\bea
%\hat{\beta}_{10} (z) &=& \frac{2}{z^2} \Phi_0^2 (z) +
%\eta \left[  \left( \frac{2}{z^2} -2\mu_0^2 z^2 + (4+\mu_0^2) z^3 \right) \Phi_0^2(z) -2h_0(z)[\Phi_0'(z)]^2 \right. \nonumber\\
%&& \left. + \frac{-8-6\mu^2 z^4 + 5(4+\mu^2)z^3}{2z} \Phi_0(z)\Phi_0(z) -2h_0(z)\Phi_0(z)\Phi_0''(z) \right] \, .
%\eea
The mode function  $\beta_{10}$ also contributes at
$\mathcal{O} (z^3)$ near the boundary, because $\beta_{10} \sim
z^3$ at the boundary.

Finally, the mode function $h_{10}(z)$ is given by
\begin{equation}
h_{10}(z)=-\frac{\mu_0 [ 2C +\mu_0 \alpha_{10}(1)]}{4} (1-z)  - \frac{1}{4}
\int_z^1 dw\, w^{2\Delta-4} \mathcal{H} (w) \label{eq6}~, \ee
where
\bea\label{eqH}
\mathcal{H} (z) &=&\left[ m^2 +\frac{q^2 \mu_0^2 z^2 (1-z)^2}{h_0(z)}+ k^2 z^2(1+ \eta z^4 \mu_0^2)+\Delta^2 h_0(z) \right] F^2(z) \nonumber\\
&&+2 z \Delta h_0(z) F(z) F'(z)
+z^2 h_0(z)[F'(z)]^2 \\
\nonumber
&& - z^{4-2\Delta} \mu_0^2 \int_1^z dw w^{2\Delta}F(w) \left[\left(\frac{2q^2(1-w)}{w^2h_0(w)}+4 \tau \eta (\Delta+1) w \right)F(w)+4\tau \eta w^2 F'(w)\right]
~.
\eea
%\nonumber\\
%&& -4z h_0^2(z) \Phi_0'(z)\Phi_0''(z) -\frac{k^2z(16+8\mu_0^2 z^4 -7(4+\mu_0^2)z^3)}{4}\Phi_0(z)\Phi_0'(z)\nonumber\\
%&& -2k^2 z^2h_0(z)\Phi_0(z)\Phi_0''(z) \Big]
The mode function $h_{10}$ contributes at $\mathcal{O}
(z^3)$ to the  metric  (\ref{ansatz}) near the boundary because $h_{10} (0)$ is finite, and we removed a factor of $z^3$ in the definition
(\ref{fourier}). We fix one of the integration constants by setting $h_{10} (1)=0$, so that the horizon
remains at $z=1$. $C$ (eq.\ \eqref{eqC}) is the remaining integration constant to be
determined.

The remaining first-order modes $\alpha_{11}, \beta_{11}, h_{11}, A_{t11}$ are determined by a system of coupled linear ordinary differential equations,
\bea
\alpha_{11}'+\frac{zh_0'+3h_0+2k^2 z^2}{zh_0} \alpha_{11}-\frac{4k^2 z}{h_0} h_{11}- \frac{z^{2\Delta -4}}{2 h_0} \mathcal{A}_1 &=& 0~,\nonumber\\
\beta_{11}'-\frac{3}{2} z \mu_0  A_{t11}'-4h_0 h_{11}'+3\frac{2 k^2 z^2+h_0 }{ z h_0}\beta_{11}-\frac{\mu_0  \left( 5h_0 +3zh_0'\right) }{2h_0}A_{t11} && \nonumber\\
+\frac{1}{4} \left(-8 k^2z+3\mu_0^2 z^3 +2h_0'\right)\alpha_{11}+\frac{\left(10 k^2 z^2 -h_0 -8z h_0'\right) }{z}h_{11}+ \frac{z^{2\Delta -4}}{4 h_0^2} \mathcal{A}_2 &=& 0~,\nonumber\\
h_{11}'-\alpha_{11}'-\frac{ \beta_{11}'}{h_0}+\left[ \frac{1}{z} + \frac{2h_0'}{h_0} \right]h_{11}+\frac{ \mu_0  A_{t11}}{h_0}
-\frac{3}{2} \left[ \frac{2}{z} + \frac{h_0'}{h_0} \right] \alpha_{11}-\frac{3\beta_{11}}{z h_0}+ \frac{z^{2\Delta -4}}{2h_0} \mathcal{A}_3&=& 0~, \nonumber\\
A_{t11}''+2 \left[ \frac{1}{z} + \frac{h_0'}{h_0} \right] A_{t11}'+\frac{2h_0'+z(-4k^2+h_0'')}{zh_0}A_{t11}
-\frac{\mu_0}{2} z^2  \alpha_{11}'&&\nonumber\\
-\frac{ \mu_0}{2} z^2  \left[ \frac{3}{z} + \frac{h_0'}{h_0} \right] \alpha_{11}-\frac{z^{2\Delta-3}\mu_0 F^2\left[q^2(1-z)-2 \tau \eta  z^3 h_0\right]}{h_0^2}&=& 0~,~~~~
\label{eqV11}
\eea
where
%Functions $A(z)$, $B(z)$, and $C(z)$ are given below.
\bea\label{A1}
\mathcal{A}_1(z)&=&z^2 h_0^2 {F'}^2+2 z \Delta h_0^2 FF'+\left[q^2 (1-z)^2 \mu_0^2+\Delta^2 h_0^2\right] F^2~,
\nonumber\\
\mathcal{A}_2(z)&=&5 z^2 h_0^2 {F'}^2+2 z (5 \Delta-1) h_0^2 F F'\nonumber\\
&& +\left[5q^2 \mu_0^2 (1-z)^2 z^2+ 3 \left(m^2-k^2 z^2 (1+2\eta z^4 \mu_0^2)\right)h_0+\Delta (5\Delta-2) h_0^2\right] F^2 ~,
\nonumber\\
\mathcal{A}_3(z)&=&F \left(\Delta F+z F'\right) \label{A3}, \eea
with $F$ defined as in (\ref{Fscalar}). The system of
equations \eqref{eqV11} can be seen to possess a unique
solution by requiring finiteness of all functions in the entire
domain $z\in [0,1]$. Notice that the unknown parameter
$\mathcal{C}$ is absent, which is due  to the fact that at
first-order the $10$ modes decouple from the $11$ modes (see
eq.\eqref{fourier} for the definition of the Fourier modes).
However, explicit solutions can only be obtained numerically.
%; see Fig. \ref{fig11} for an example.
A complete numerical analysis
will be presented elsewhere.

%\begin{figure}
%\begin{center}
%{\includegraphics[width=.4\textwidth]{alpha11.eps}}$~~~~~~~~~~$
%{\includegraphics[width=.4\textwidth]{h11.eps}}
%{\includegraphics[width=.4\textwidth]{beta11.eps}}$~~~~~~~~~~$
%{\includegraphics[width=.4\textwidth]{At11.eps}}
%\end{center}
%\caption {Profiles of the Fourier modes $\alpha_{11}$ (upper left panel), $h_{11}$ (upper right panel), $\beta_{11}$ (lower left panel), and $A_{t11}$ (lower right panel) for $\eta/{\mu^2} =0.3$,  $\eta'/{\mu^4} =0.005$, $\Delta =1.51$, $q=1.17$.
%} \label{fig11}
%\end{figure}
%

To complete the determination of the first order modes, we need to calculate the integration constant $C$ (or, equivalently, the chemical potential parameter $\mu_1$ - see eq.\ \eqref{eqC}).
To this end, we turn to the scalar wave equation. At zeroth order, the chemical potential parameter $\mu_0$ was obtained as an eigenvalue of the scalar wave equation. The first-order correction, $\mu_1$, is determined by
the first-order equation of the scalar wave equation.

Considering \eqref{waveeq} below the critical temperature,  the scalar field at first order has two Fourier modes,
% can be written as
%, we make a Fourier series expansion of the scalar field. At first order, we only have two modes,
\be
\phi_1(z,x)=\Phi_{10}(z) \cos kx+\Phi_{11}(z) \cos 3kx~. \ee
The first ($\Phi_{10}$) mode satisfies the equation
%We deduce from (\ref{waveeq})~,
\be
\Phi_{10}''+\left[\frac{h_0'}{h_0}-\frac{2}{z}\right] \Phi_{10}'+\frac{\tau}{h_0}\left(1- \eta \mu_0^2 z^4\right)\Phi_{10}-\frac{1}{h_0} \left[\frac{m^2}{z^2}-q^2\frac{A_{t0}}{h_0}\right] \Phi_{10} + z^{\Delta+1}\mathcal{B} + Cz^{\Delta+2}\mathcal{C} =0~,\ee
where
\be
\mathcal{B} = \mathcal{B}_2 F''+\mathcal{B}_1 F' + \mathcal{B}_0 F \ \ , \ \ \ \ \mathcal{C} =\mathcal{C}_2 F''+\mathcal{C}_1 F' + \mathcal{C}_0 F~,
\label{newweq}
\ee
and %$f_{\pm} = 1 - \eta\frac{\mu_0^2}{4} z^4$, and %
the coefficients $\mathcal{B}_i$ and $\mathcal{C}_i$ ($i=0,1,2$) are
\bea
\mathcal{B}_0&=&\frac{z \mu_0^2 \left[q^2 (1-z)^2+k^2 z^4 \eta  h_0\right]}{h_0}\alpha_{10}-\frac{1}{2} \Delta  h_0 \alpha_{10}'+\tau z (1-\eta z^4 \mu_0^2)\beta_{10}\nonumber\\
&&+\frac{z^{2 } \left[-q^2 (1-z)^2 z^2 \mu_0^2+\Delta ^2 h_0^2\right] }{h_0^2}\mathbf{h}+z^{3 } \Delta \mathbf{h}'-\frac{2 q^2 (-1+z) z \mu_0 }{h_0}\mathbf{A}\nonumber\\
&&-2\tau \eta \mu_0 z^{5} \mathbf{A}'+\frac{1}{4} z^{2 }  \left(2 q^2 (1-z)^2 z^2 \mu_0^2-3 \Delta  h_0^2-z h_0 \left(2 k^2 z+\Delta  h_0'\right)\right)\alpha_{11}\nonumber\\
&&-\frac{1}{4} z^{3 } \Delta  h_0^2 \alpha_{11}'-\frac{1}{2}\tau z^{4} \left(1-\eta z^4 \mu_0^2\right)\beta_{11}-z^{2 } \mu_0 \left(q^2 (-1+z)+2 k^2 z^3 \eta h_0\right)A_{t11}\nonumber\\
&&+\frac{1}{2} z^{2 }  \left(-q^2 (1-z)^2 z^2\mu_0^2+\Delta ^2 h_0^2+2 z \Delta  h_0 h_0'\right)h_{11} +\frac{1}{2} z^{3 } \Delta  h_0^2 h_{11}' ~,\nonumber\\
\mathcal{B}_1&=&z^{3 } (1+2 \Delta ) \mathbf{h}+z^{4 } \mathbf{h}'-\frac{1}{2} zh_0 \alpha_{10}'-\frac{1}{4} z^{3 } h_0  \left(3 h_0+z h_0'\right)\alpha_{11}-\frac{1}{4} z^{4 } h_0^2 \alpha_{11}'\nonumber\\
&&+\frac{1}{2} z^{3 } h_0 \left(h_0+2 \Delta  h_0+2 z h_0'\right)h_{11}+\frac{1}{2} z^{4 } h_0^2 h_{11}' ~,\nonumber\\
\mathcal{B}_2&=& z^{4} \mathbf{h}+\frac{1}{2} z^{4} h_0^2 h_{11}~,
\eea
\bea
\mathcal{C}_0&=&\frac{\mu_0 }{2 h_0^2}\left[q^2 (1-z)^2\left((z-1)z^3\mu_0^2+4h_0\right)+z\left(-\Delta^2+ \Delta (\Delta+1)z+4 \tau \eta z^3\right)h_0^2\right] ~,
\nonumber\\
\mathcal{C}_1&=&\frac{\mu_0}{2} z^{2 } \left[-1-2 \Delta +2 z (1+\Delta )\right] ~,
\nonumber\\
\mathcal{C}_2&=&\frac{\mu_0}{2} (z-1) z^{3 } ~.
\eea
We defined (see eqs.\ \eqref{eqC} and \eqref{eqH})
\bea
\mathbf{h}(z)&=& - \frac{1}{4} \int_z^1 dw\, w^{2\Delta-4} \mathcal{H} (w) ~,\\
\nonumber
\mathbf{A}(z)&=&\frac{\mu_0}{4} \int_1^z dw\, \int_1^w dw' {w'}^{2\Delta-2} h_0(w') \mathcal{A}(w')~.
\eea
By using the zeroth order wave equation \eqref{weqPsi}, we obtain
% the eigenvalue
\be C=-\frac{ \int_0^1 dz \, z^{2\Delta+1} F \left[ \mathcal{B}_2 F''+\mathcal{B}_1 F' + \mathcal{B}_0 F\right]}{\int_0^1 dz\, z^{2\Delta+2}  F\left[\mathcal{C}_2 F''+\mathcal{C}_1 F' + \mathcal{C}_0 F\right]}~. \ee
Having obtained the integration constan $C$, the remaining unknown parameter $\mu_1$ is calculated using eq.\ \eqref{eqV10}.

The temperature of our
system below the critical temperature $T_c$ can be calculated using
\be \frac{T}{\mu} = - \frac{h'(1) e^{-\alpha(1)}}{4\pi\mu}~. \ee
We obtain
\be\label{eq2_38} \frac{T}{T_c} = 1-\xi^2 \left( \alpha_{10}(1) + \frac{\mu_1}{\mu_0} \right) -
\frac{\xi^2}{3- \frac{\mu_0^2}{4}} h_{10}'(1)~, \ee where $\xi$ is given by
eq. (\ref{condensate1}).

Eq.\ \eqref{eq2_38} can be inverted to find the energy gap  (\ref{condensate1}) as a
function of temperature near the critical temperature, \be
\frac{\langle\mathcal{O}_\Delta \rangle^{1/\Delta}}{T_c} \approx
\gamma \left( 1- \frac{T}{T_c} \right)^{\frac{1}{2\Delta}} \ \ , \
\ \ \ \gamma = \frac{4\pi}{3-\frac{\mu_0^2}{4}} \left( \frac{\alpha_{10} (1)}{2} +
\frac{\mu_1}{2\mu_0} + \frac{h_{10}'(1)}{2(3-\frac{\mu_0^2}{4})} \right)^{-\frac{1}{2\Delta}}~. \ee
 Thus,  as the
temperature
 of the system is lowered below the critical temperature
 $T_c$ the condensate is spontaneously generated. The dependence of the condensate on the temperature is of the same form as in conventional holographic superconductors.

Finally, the charge density of the system is determined by using
\be \frac{\rho}{\mu^2} = - \frac{\partial_z
A_t(0,x)}{[A_t(0,x)]^2} =\frac{\rho_0+ \xi^2 \rho_1(x)}{\mu_0^2}~,
\ee where $\rho_0=\mu_0$ is the charge density at or above the
critical temperature, and \be \rho_1(x) = -2\mu_1 -A_{t10}'(0)-
A_{t11}(0) \cos 2kx ~,\ee from eq.\ (\ref{eq4}). This is an important result
showing the generation  of a spatially inhomogeneous charge
density below the critical temperature in the presence of a
\emph{spatially homogeneous} (constant) chemical potential. This
is the case provided $A_{t11}(0) \ne 0$, which is guaranteed
analytically from the system of equations \eqref{eqV11} for the
$11$ modes. Indeed, from the last equation in \eqref{eqV11}, we
obtain $A_{t11}'(0)=0$. Moreover, there is a boundary condition at
the horizon $z=1$ where we demand finiteness of $A_{t11}$
($A_{t11} (1) < \infty$). If additionally $A_{t11}(0)=0$, then the
second-order differential equation is overdetermined and has no
solution. Thus, a general solution has $A_{t11} (0) \ne 0$.
% This is confirmed by numerical analysis (see Fig. \ref{fig11}).

\section{Conclusions}
\label{sec5}

We have discussed a holographic model in which the gravity sector
consists of a $U(1)$ gauge field and a scalar field coupled to an
AdS charged black hole under a \emph{constant} chemical potential. We introduced higher-derivative interaction terms between
the $U(1)$ gauge field and the scalar field. A
gravitational lattice was generated spontaneously by a spatially dependent profile of the
scalar field.  The transition temperature was calculated as a function of the wavenumber. The critical temperature was determined as the maximum transition temperature. This occurred at finite non-vanishing wavenumber, showing that the inhomogeneous solution was dominant over a range of the parameters of the system which we discussed.

The system was then studied below the critical temperature. We obtained analytic expressions for the various fields using perturbation theory in the (small) order parameter. It was found that a spatial
inhomogeneous phase is generated at the boundary. In particular, we showed analytically that a spatially inhomogeneous charge density is spontaneously generated in the system while it is held at constant chemical potential.

It will be illuminating to compare features between different mechanisms for generating spontaneous translation symmetry breaking seen with the Einstein tensor-scalar coupling  \cite{Alsup:2012kr}, Chern-Simons interaction \cite{Donos:2012cs,Nakamura:2009}, and dilaton \cite{Donos:2013bh}.  Additionally, this work only considered a uni-directional lattice.
It would be interesting to extend our discussion to a more general
two-dimensional lattice and determine which configuration is energetically favorable.  Finally, one would like to understand the origin of the higher-derivative couplings we introduced in terms of quantum corrections within string theory. Work in these directions is in progress.

\acknowledgments
G.\ S.\ and K.\ Y.\ were supported in part by the US Department of Energy
under Grant No.\ DE-FG05-91ER40627.

\end{document}